\documentclass[a4paper,11pt]{article}
\usepackage{jheppub}
\usepackage{multicol,enumerate}
\usepackage{multirow,amsmath,amssymb}
\usepackage{subfigure}
\usepackage{slashed} 
\usepackage{color} 

\newcommand{\whizard}{{\sc Whizard}}
\newcommand{\madgraph}{{\sc MadGraph}}

\title{Initial State Radiation Simulation with MadGraph}

\author{Qiang Li$^{a}$,}
\author{Qi-Shu Yan$^{b}$}

\affiliation{$^a$Department of Physics and State Key Laboratory of Nuclear Physics and Technology, \\
Peking University, Beijing, 100871, China}
\affiliation{$^b$College of Physics Sciences, University of Chinese Academy of Sciences, Beijing 100049, China and Center for High Energy Physics, Peking University, Beijing 100871, China}

\emailAdd{ qliphy0@pku.edu.cn, yanqishu@ucas.ac.cn}

\abstract{Initial State Radiation (ISR) effect is implemented in \madgraph~using photon structure functions. Detailed examinations are performed against other generators including \whizard, BabayagaNLO and KKMC, for various physics processes at linear colliders such as 240GeV CEPC and 10.58GeV Belle II.  ISR effects on new physics searches are also discussed, taking anomalous gauge boson coupling and dark photon as examples. An ISR plugin~\footnote{\url{http://github.com/qliphy/MGISR}} is made public for wide usage.}

\date{\Date}


\begin{document}
\maketitle
\flushbottom

\section{Introduction}
\label{intr}

\qquad The Initial State Radiation (ISR) is an important issue in high energy processes, especially for lepton colliders, including CEPC~\cite{ref:cepc_1,ref:cepc_2}, ILC~\cite{ref:ilc} and B-factories (for example Belle II~\cite{Abe:2010gxa,b2tip}). ISR affects both total and differential cross section significantly, for example, reduces the $ZH$ inclusive cross section at CEPC by more than 10\%~\cite{Chen:2017gzv}. We have managed for the first time to simulate ISR in \madgraph~ with lepton ISR structure function~\cite{pstruc1}, that includes all orders of soft and soft-collinear photons as well as up to the third order in hard-collinear photons. Comparisons can be seen in Ref.~\cite{Chen:2017gzv} for  $e^+e^- \rightarrow ZH$, from which one can see the good agreement between \whizard~\cite{ref:4}, and \madgraph with ISR included. 

Within \madgraph~version 26X, a plugin is now made public to further simplify ISR simulation. The plugin, on top of the user friendly framework of \madgraph, makes relevant studies at linear colliders more accurate, including searching for dark photon~\cite{Aaij:2017rft,Ablikim:2017aab,Lees:2014xha} and probing anomalous triple or quartic gauge couplings~\cite{Yang:2012vv,aqgc1}.

This paper is organized as follows. Section 2 shows the ISR implementation in \madgraph and results at CEPC . Section 3 provides results at Belle II.  Section 4 presents the ISR effects on new physics searches. The conclusion is summarized in Section 5.

\section{$e^+e^- \rightarrow W^+W^- \gamma$ at CEPC}
\label{isrmg}
\begin{figure}[!t]
  \centering
  \includegraphics[width=0.48\textwidth]{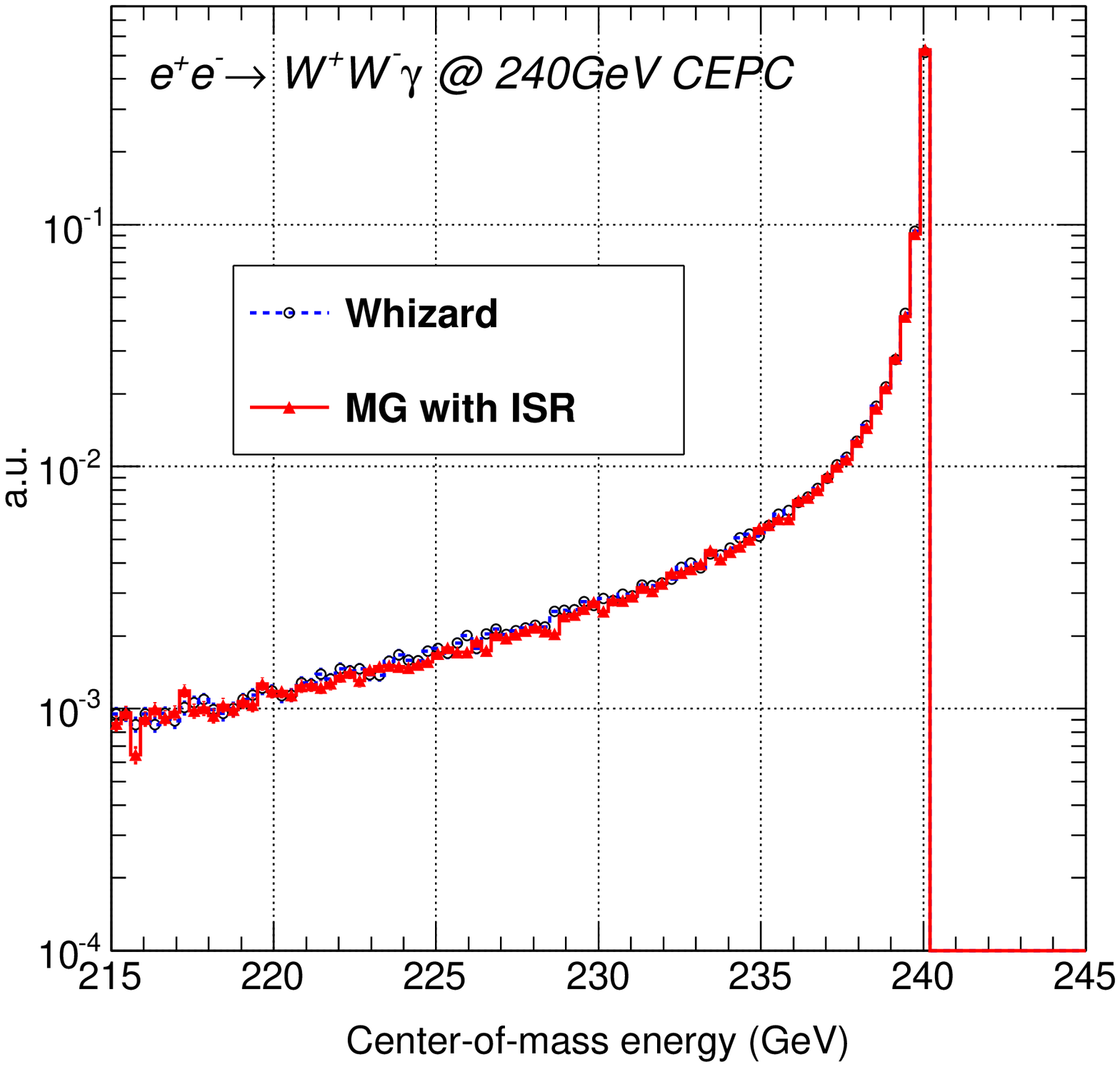}
  \includegraphics[width=0.48\textwidth]{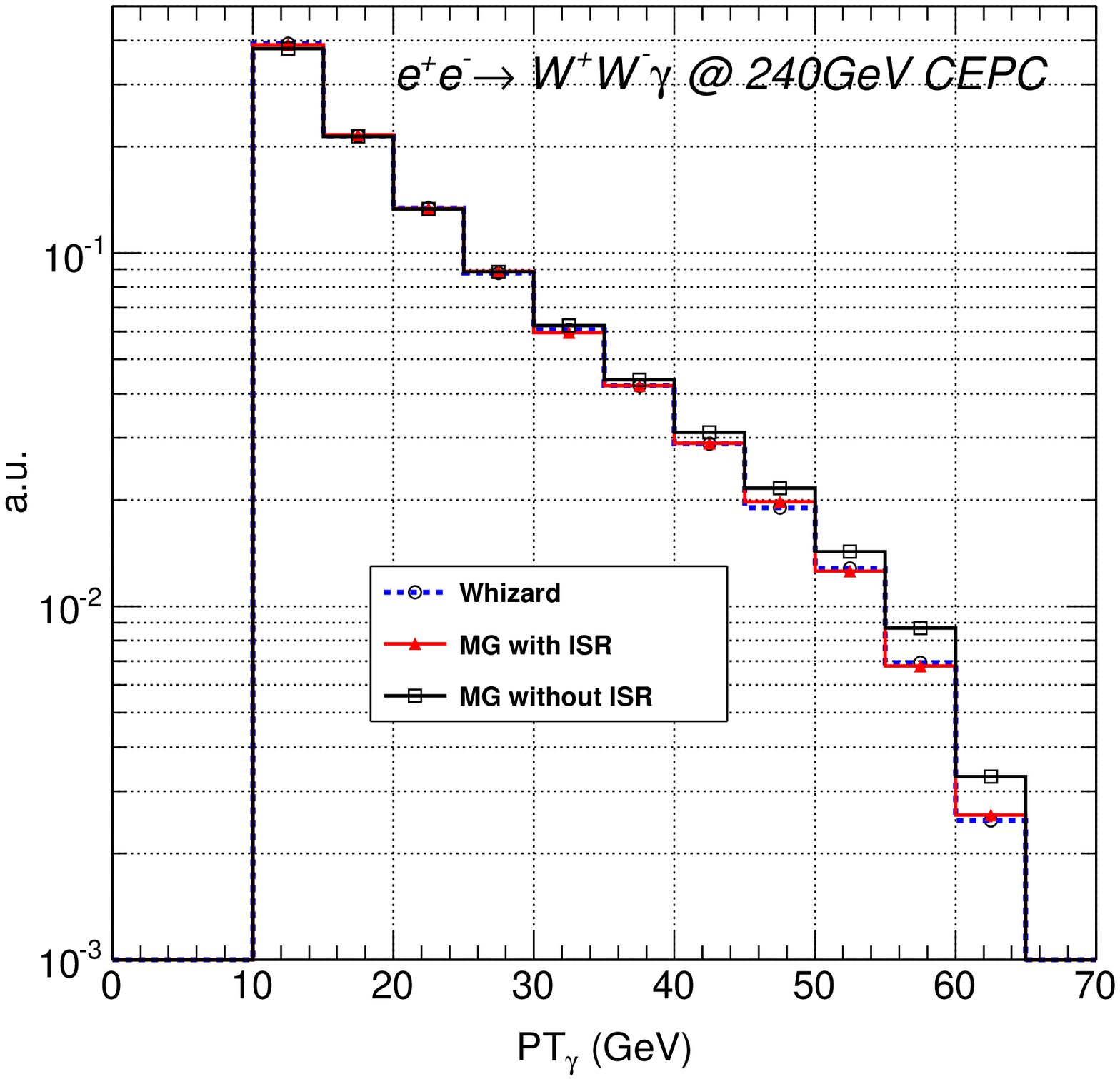}
  \caption{\label{fig:isr} Comparisons plots on center-of-mass energy and photon transverse momentum,  between \whizard~ and \madgraph~ with or without ISR effect included, for the process $e^+e^- \rightarrow W^+W^- \gamma$.}
\end{figure}

\qquad $W^+W^- \gamma$ production at CEPC and ILC is an interesting process to test the Standard Model and probe anomalous quartic gauge couplings. At 240GeV CEPC with photon transverse momentum $\rm{PT}_{\gamma}>10$~GeV, the cross sections from \madgraph~read  263.2 and 237.4fb, without or with ISR included, respectively. Fig.~\ref{fig:isr} gives the normalized differential distributions for $e^+e^- \rightarrow W^+W^- \gamma$ at 240GeV CEPC, from which one can see the good agreement between \whizard~and \madgraph~with ISR included, on distributions of center-of-mass energy and photon transverse momentum. With ISR included, $\rm{PT}_{\gamma}$ tends to be softer as expected.

\section{$e^+e^- \rightarrow \mu^+\mu^-$ at Belle II}
\label{isrmg}
\begin{figure}[!t]
  \centering
  \includegraphics[width=0.48\textwidth]{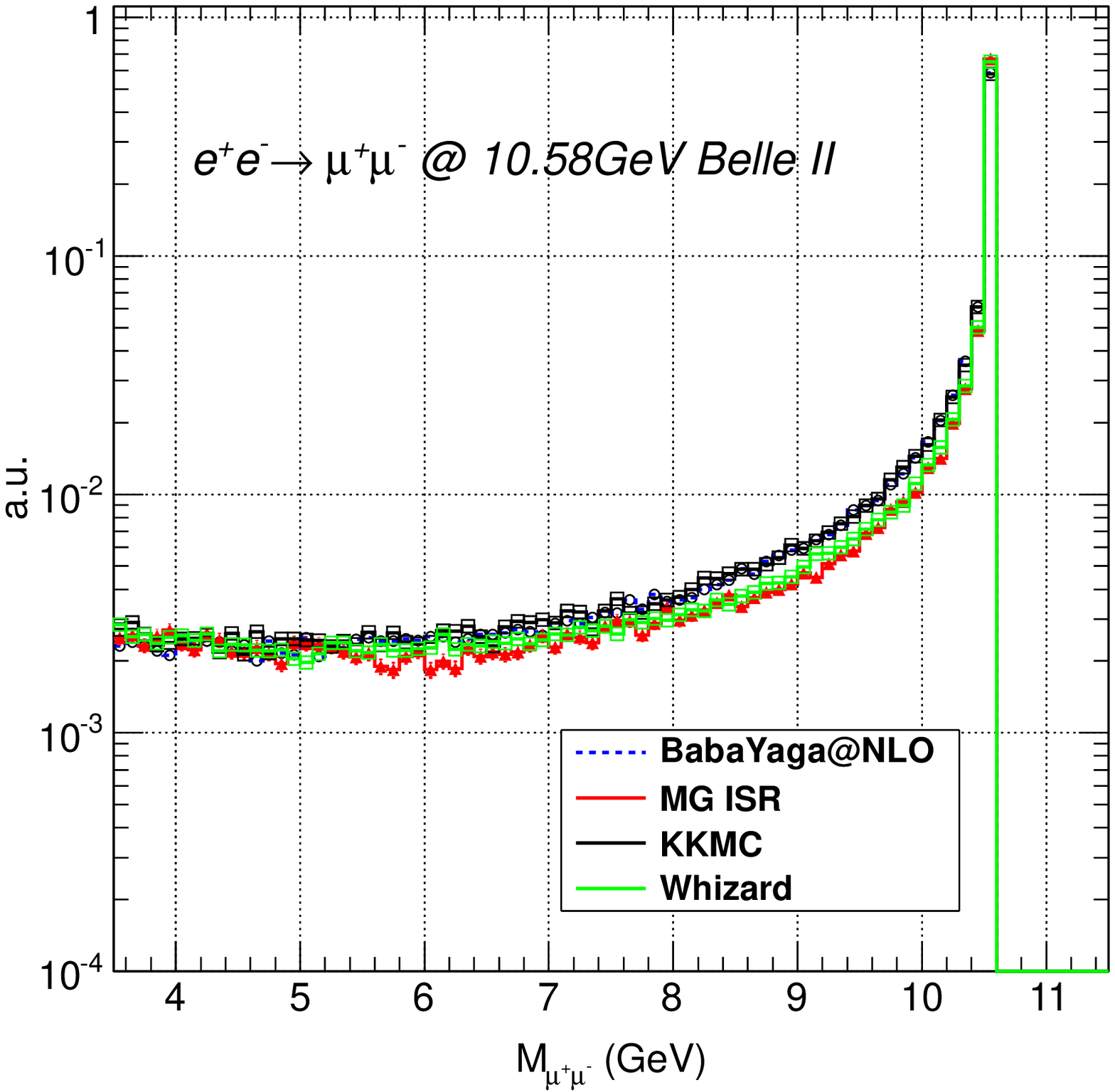}
  \includegraphics[width=0.48\textwidth]{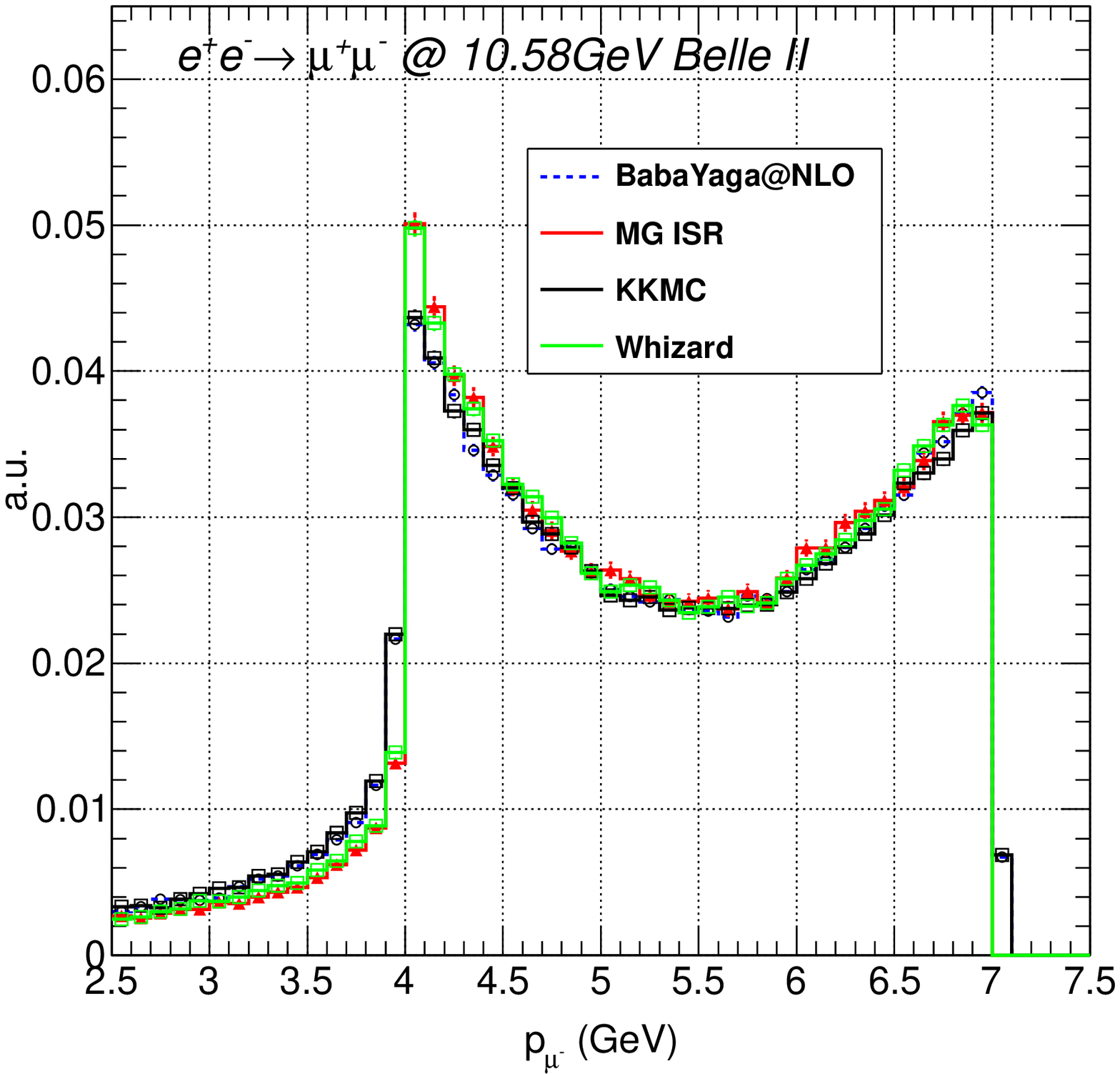}
  \caption{\label{fig:isr2} Comparisons plots on muon pair invariant mass and muon momentum, among various generators, for the process $e^+e^- \rightarrow \mu^+\mu^-$ at Belle II (4GeV position + 7 GeV electron as beams). All the curves are normalized to same area. }
\end{figure}

\qquad $e^+e^- \rightarrow \mu^+\mu^-$ at Belle II is crucial for background control and luminosity measurement. The cross section from \madgraph~with ISR reads 1.14nb (\madgraph~without ISR gives about 0.86nb), which agrees well with the precision prediction from BabayagaNLO~\cite{babayaganlo} and KKMC~\cite{kkmc}. Fig.~\ref{fig:isr2} provides further detailed comparison among \whizard, BabayagaNLO, KKMC and \madgraph. In general, good agreement are found between \whizard~and \madgraph, and between KKMC and BabayagaNLO. The discrepencies between the former and later should be related to different treatment of photon structure functions and QED coupling running, etc. Various simulations will be useful for future systematic studies.

\section{ISR effects on aQGC and Dark photon searches}
\label{results}

\begin{figure}[!t]
  \centering
  \includegraphics[width=0.78\textwidth]{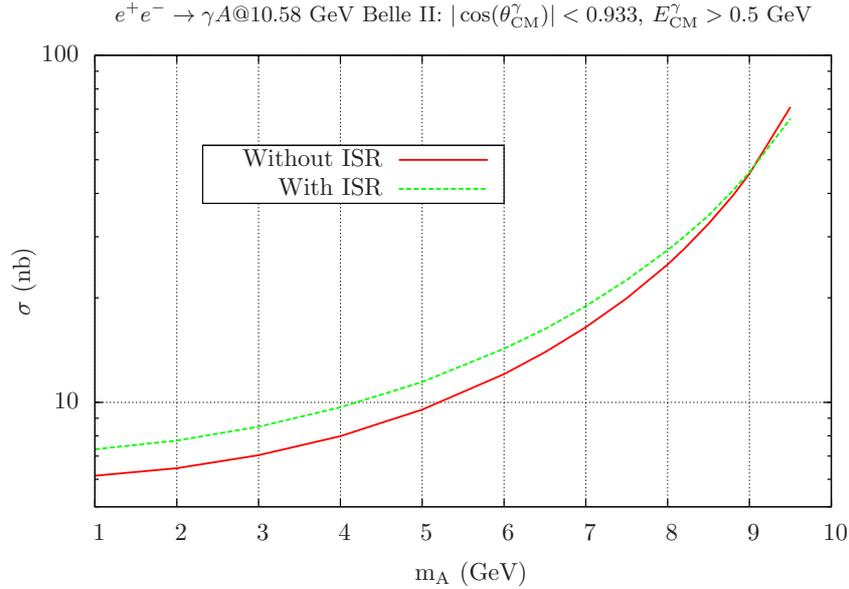}
  \caption{\label{fig:isr3} Cross section dependence on dark photon mass, for $e^+e^- \rightarrow \gamma A$ at 10.58 GeV Belle II, with cuts $|\cos(\theta^\gamma_{\rm CM})|<0.933$ and $E_{\rm CM}^{\gamma}>0.5$~GeV, with or without ISR effect included.}
\end{figure}

\begin{figure}[!t]
  \centering
  \includegraphics[width=0.68\textwidth]{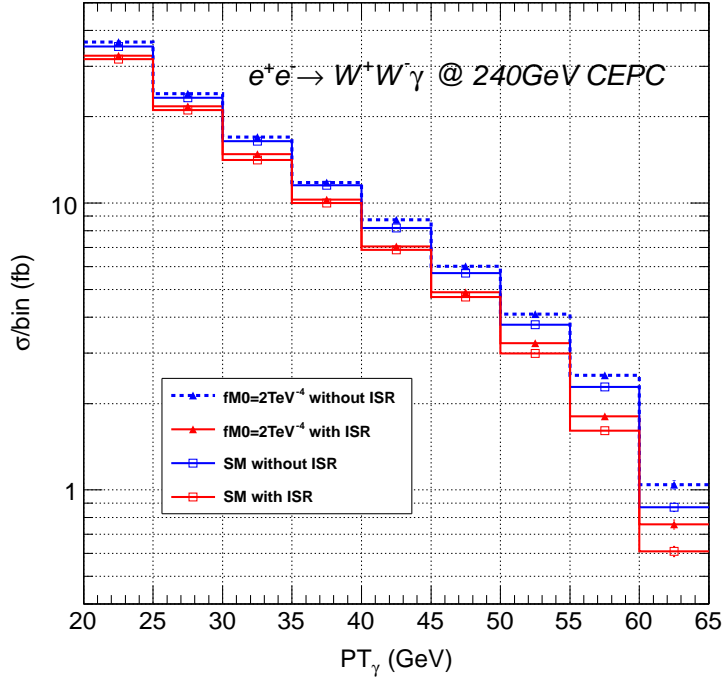}
  \caption{\label{fig:isr4} Comparisons plots on photon transverse momentum, in SM or aQGC (fM0=2~$\rm{TeV}^{-4}$), with or without ISR effect included, for the process $e^+e^- \rightarrow W^+W^- \gamma$.}
\end{figure}

\qquad As mentioned above, the ISR plugin on top of the user friendly framework of \madgraph, can make new physics simulations at linear colliders more accurate, including searching for dark photon~\cite{Aaij:2017rft,Ablikim:2017aab,Lees:2014xha} and probing anomalous triple or quartic gauge couplings~\cite{Yang:2012vv,aqgc1}. Fig.~\ref{fig:isr3} shows the dark photon production rate dependence on dark photon mass, for $e^+e^- \rightarrow \gamma A$ at 10.58 GeV Belle II, with cuts $|\cos(\theta^\gamma_{\rm CM})|<0.933$ and $E_{\rm CM}^{\gamma}>0.5$~GeV, with or without ISR effect included. For light dark photon, the cross section with ISR is larger by 10-20\% than the one without ISR included. Fig.~\ref{fig:isr4} shows the comparisons on photon transverse momentum,  in SM or aQGC (fM0=2~$\rm{TeV}^{-4}$), with or without ISR effect included, for the process $e^+e^- \rightarrow W^+W^- \gamma$. One can see the ISR is crucial to be included for aQGC searches, as it changes the tail distributions largely. 

\section{Summary and Conclusions}
\label{talk}
\qquad ISR effect has been implemented in \madgraph~ and a easy-to-use plugin is provided. Detailed examinations are perfomed against other generators including \whizard, BabayagaNLO and KKMC, for various physics processes at linear colliders such as 240GeV CEPC and 10.58GeV Belle II. ISR effects on new physics searches are also discussed, taking anomalous gauge boson coupling and dark photon as examples, and found to be crucial.

\acknowledgments
We thank Dr. Olivier Mattelaer and Dr. Carlo Carloni Calame for valuable help. This work is supported in part by the National Natural Science Foundation of China, under Grants No. 11475190 and No. 11575005,  by the CAS Center for Excellence in Particle Physics (CCEPP).

\appendix


\end{document}